\documentclass[9pt,twocolumn]{extarticle}

\usepackage{titlesec}
\titlespacing*{\section}
{0pt}{0.3cm}{0.3cm}

\titlespacing*{\subsection}
{0pt}{0.2cm}{0.2cm}

\font\myfont=cmr12 at 20pt
\newcommand{\secfnt}{\fontsize{11}{11}}
\newcommand{\ssecfnt}{\fontsize{10}{10}}

\titleformat{\title}
{\tfnt\bfseries}{\thetitle}{1em}{}

\titleformat{\section}
{\normalfont\secfnt\bfseries}{\thesection}{1em}{}
 
\titleformat{\subsection}
{\normalfont\ssecfnt\bfseries}{\thesubsection}{1em}{}

\usepackage[left=1.7cm, right=1.5cm, top=1.5cm, bottom=1.7cm]{geometry}
\usepackage{amsmath}
\usepackage{multirow}
\usepackage{latexsym}
\usepackage{subfigure} 
\usepackage{amssymb}
\usepackage{graphicx}
\usepackage{epstopdf}
\usepackage{color}
\usepackage{xcolor}
\usepackage[toc,page]{appendix}
\usepackage{algorithm}
\usepackage{algorithmic}
\usepackage{lipsum}
\usepackage{fancyhdr}
\usepackage{authblk}

\usepackage{amsthm}

\definecolor{upmaroon}{rgb}{0.69, 0.11, 0.23}

\usepackage{graphicx}
\date{\vspace{-4ex}} 
\newtheorem{definition}{Definition}

\begin{document}
%
%
\title{
\myfont Positive Geometries for Barycentric Interpolation
}
\author[1]{
Márton Vaitkus
}
\affil[1]{Budapest University of Technology and Economics}
\setcounter{Maxaffil}{0}
\renewcommand\Affilfont{\itshape\small}

\maketitle
%
%

\textit{
\begin{abstract}
We propose a novel theoretical framework for barycentric interpolation, using concepts recently developed in mathematical physics. Generalized barycentric coordinates are defined similarly to Shepard's method, using positive geometries -- subsets which possess a rational function naturally associated to their boundaries. Positive geometries generalize certain properties of simplices and convex polytopes to a large variety of geometric objects. Our framework unifies several previous constructions, including the definition of Wachspress coordinates over polytopes in terms of adjoints and dual polytopes. We also discuss potential applications to interpolation in 3D line space, mean-value coordinates and  splines. 
\end{abstract}
}
%
%
%
%

\section{Overview}
\label{sec:intro}
The interpolation of scalar- or vector-valued data is an important task in many fields, including numerical analysis, geometric modeling and computer graphics. \emph{Barycentric coordinates} can be defined over segments, triangles and simplices as well as more complex shapes, such as polytopes  \cite{floater2015generalized}. These include \emph{Wachspress coordinates} \cite{wachspress2016rational} which are rational functions defined for convex polytopes. Wachspress coordinates were also generalized to subsets bounded by non-linear hypersurfaces \cite{dasgupta2008adjoint} and can be defined in several equivalent ways \cite{warren1996barycentric,meyer2002generalized,ju2005geometric}. We describe a theoretical framework that clarifies the relationship between these different formulations, and provides opportunities for novel generalizations.

We propose a set of basic building blocks for barycentric interpolation methods: \emph{positive geometries}. Positive geometries have been recently introduced in the theoretical physics literature \cite{arkani2017positive}, but their application to interpolation problems has not been explored before. Besides simplices and polytopes, the category of positive geometries includes objects with similar combinatorial properties, such as polycons \cite{wachspress2016rational} or ``positive" parts of toric varieties \cite{sottile2003toric} and Grassmannians \cite{pottmann2001computational,gelfand1982geometry}. A positive geometry carries a ``canonical" differential form: a rational function with the properties of a signed volume, that has its poles (where the denominator vanishes) along the boundaries of the ``positive" region. Crucially, these poles have a recursive structure, i.e. restricting a canonical form to a boundary component (via a generalization of taking complex residues) results in another canonical form. Thus, positive geometries share many properties with polytopes -- most importantly, that complicated objects can be constructed by adding together simpler ones. To define interpolation in terms of canonical forms, we use a variant of Shepard's method \cite{floater2015generalized,rvachev2001transfinite}: barycentric coordinates are ratios tending to $\frac{\infty}{\infty}$ along the interpolated boundaries. For polytopes, our construction recovers the definition of Wachspress coordinates in terms of dual volumes \cite{ju2005geometric}.

In this preliminary work, our goal is to introduce the theory of positive geometries to our audience, and demonstrate how they generalize earlier constructions for barycentric interpolation. After some motivating observations (Section 2), we give a definition of positive geometries and their canonical forms, also giving some examples (Section 3). We then describe how to use the canonical forms of positive geometries for barycentric interpolation (Section 4). Finally, we discuss potential applications of this framework to interpolation in line space, and possible extensions to non-rational barycentric coordinates and splines (Section 5).

\section{Motivation}

Let us consider the basic example of linear interpolation over a segment $[a,b]\subset \mathbb{R}$. The standard formula 
\begin{align}
f(x) = \frac{b - x}{b - a}f(a) + \frac{x - a}{b - a}f(b),
\end{align}
can be written in an alternative \emph{barycentric form} \cite{hormann2014barycentric}:
\begin{align}\label{eq:segment_bary}
f(x) = \frac{\frac{1}{x - a} f(a) - \frac{1}{x - b}f(b)}{-\frac{b - a}{(x - a)(x - b)}},
\end{align}
as a rational function with both the numerator and the denominator having poles at the boundaries $a$ and $b$.

\begin{figure}[h]
	\centering
	\includegraphics[width=0.5\textwidth]{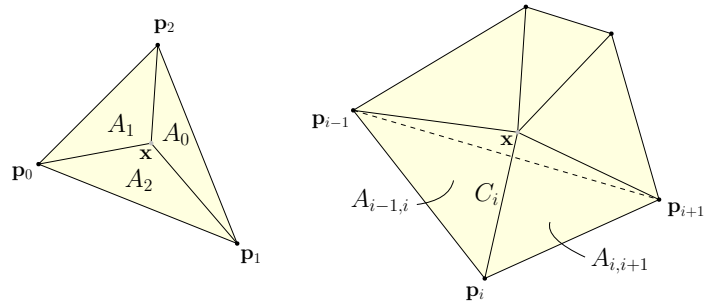}
	\caption{Notations for barycentric coordinates.}
	\label{fig:bary_notation}
\end{figure}

Consider next linear interpolation over triangles $(\mathbf{p}_0,\mathbf{p}_1,\mathbf{p}_2) \subset \mathbb{R}^{2}$ in the plane. The usual formula for barycentric interpolation gives
\begin{align}
f(\mathbf{x}) = \frac{A_{0}f(\mathbf{p}_0) + A_{1}f(\mathbf{p}_{1}) + A_2 f(\mathbf{p}_2)}{A}, 
\end{align} 
which can be reorganized into the equivalent form
\begin{align}\label{eq:triangle_bary}
f(\mathbf{x}) = \frac{\frac{1}{A_{1}A_{2}}f(\mathbf{p}_0) + \frac{1}{A_{2}A_{0}}f(\mathbf{p}_1) + \frac{1}{A_{0}A_{1}}f(\mathbf{p}_2)}{\frac{A}{A_{0}A_{1}A_{2}}},
\end{align}
with the notations shown in Figure \ref{fig:bary_notation}.

\begin{figure}[h]
	\centering
	\includegraphics[width=0.4\textwidth]{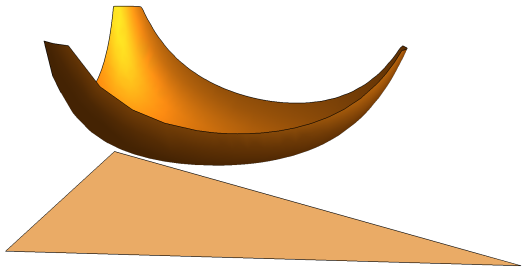}
	\caption{Canonical rational function of a triangle.}
	\label{fig:triangle_form}
\end{figure}

Wachspress proposed generalized barycentric coordinates over convex polytopes \cite{wachspress2016rational}. For a planar $n$-gon with vertices $\mathbf{p}_{0}, \ldots \mathbf{p}_{n-1}$ these coordinates are defined as:
\begin{align}\label{eq:polygon_bary}
f(\mathbf{x}) = \sum_{i = 0}^{n} \frac{\frac{C_{i}}{A_{i-1,i}A_{i,i+1}}}{\frac{\sum_{k} C_{k}A_{k-3,k-2} \cdots A_{k+1,k+2}}{\prod_{k}A_{k,k+1} } }f(\mathbf{p}_{i}),
\end{align}
where the indices are cyclical modulo $n$ -- see Figure \ref{fig:bary_notation}. 

\begin{figure}[h]
	\centering
	\includegraphics[width=0.4\textwidth]{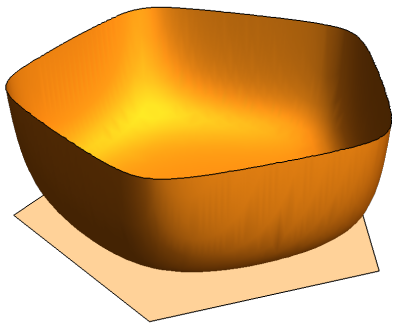}
	\caption{Canonical rational function of a pentagon.}
	\label{fig:pentagon_form}
\end{figure}
%

In each of these cases, some terms in the numerator and denominator approach infinity along an interpolated subset, which is reminiscent of Shepard's method \cite{hormann2014barycentric,rvachev2001transfinite}. The rational functions in the denominators are illustrated in Figure \ref{fig:triangle_form} and Figure \ref{fig:pentagon_form}. The common pattern involves functions with poles along the boundaries of some shape. Our goal is to make this idea rigorous using the notion of a positive geometry.

\section{Positive Geometries and Canonical Differential Forms}
The central concept of our work is that of a positive geometry -- a relatively new concept originating from mathematical physics. In this chapter, we give an informal overview, and refer to \cite{arkani2017positive} for technical details. For the necessary background in projective and algebraic geometry, see e.g. \cite{pottmann2001computational,griffiths1978principles}.

\subsection{Definition of Positive Geometries}
Positive geometries were introduced as generalizations of shapes with a recursive structure similar to polytopes, defined by some sort of ``positivity" criterion (e.g. the interior of polytopes, or totally positive matrices \cite{gasca1996total}). The definition was motivated by recent developments in quantum physics, where the solution of differential equations in Fourier space led to rational functions with poles at the boundaries of certain regions \cite{arkani2017positive}.
\begin{definition}
	A \textbf{positive geometry} is a pair $(X,X_{\geq 0})$, where $X$ is a $d$-dimensional complex projective algebraic variety, and $X_{\geq0}$ is an oriented semi-algebraic subset of its real part, with a unique differential $d$-form $\Omega(X,X_{\geq0})$ called its \textbf{canonical form}, defined by recursion on the boundary dimension:
	\begin{itemize}
		\item If $d = 0$, then $X$ is an (oriented) point, and $\Omega(X,X_{\geq0}) = \pm 1$ depending on the orientation.
		\item If $d >0$, then the boundary components of $X_{\geq 0}$ are themselves positive geometries, and the multivariate residue (see \cite[A.3]{arkani2017positive} or \cite[Ch. 5]{griffiths1978principles}) along a component is the canonical form for the associated positive geometry. 
	\end{itemize}
\end{definition} 
We stress that a positive geometry is determined by an ambient complex manifold (most often a projective space $\mathbb{P}^{n}$), together with a ``positive" real subset, thus many different positive geometries could be associated to the same ambient space.

For positive geometries over projective spaces with homogeneous coordinates $\mathbf{x} = \left[ x_{0} : x_{1} : \ldots x_{d} \right] $, the canonical form can always be written in terms of a rational function \cite[C.1]{arkani2017positive}:
\begin{align}
\Omega(X,X_{\geq0}) = C(\mathbf{x})\omega(\mathbf{x}),
\end{align}
where $C$ is called the \textbf{canonical rational function} of the positive geometry and 
\begin{align*}
\omega(\mathbf{x}) = \frac{1}{d!}x_{0}dx_{1} \wedge \ldots \wedge dx_{d} + \ldots + (-1)^{d}x_d dx_{0} \wedge \ldots \wedge dx_{d-1}
\end{align*}
is the \emph{standard measure} on projective $d$-space.

\subsection{Examples of Positive Geometries}
Some elementary examples of positive geometries are the following:
\begin{itemize}
	\item \textbf{Segments}, bounded by two points $\mathbf{a} = \left[ a:1 \right]$ and $\mathbf{b} = \left[ b:1 \right]$ in the projective line $\mathbb{P}^1 $ parameterized using homogeneous coordinates. The canonical form at the point $\mathbf{x} = \left[ x:y \right]$ is
	\begin{align}
	\Omega(\mathbb{P}^1,\left[ a,b \right]) = \frac{\left\langle \mathbf{b},\mathbf{a}\right\rangle }{\left\langle \mathbf{x},\mathbf{a}\right\rangle \left\langle \mathbf{x},\mathbf{b}\right\rangle } \omega(\mathbf{x}) ,
	\end{align}
	where $\left\langle \mathbf{v}, \mathbf{w} \right\rangle := \det(\mathbf{v}, \mathbf{w}) $ denotes the determinant of vectors of homogeneous coordinates. 
	\item \textbf{Simplices} in projective spaces. For triangles formed by three points $\mathbf{p}_{i} = \left[ x_i:y_i:1 \right], i = 0, 1, 2$ in the projective plane $\Delta \subset \mathbb{P}^2$ parameterized using homogeneous coordinates $\mathbf{x} = \left[ x:y:z \right]$, the canonical form is
	\begin{align}
	\Omega(\mathbb{P}^2, \Delta) &= \frac{1}{2}\frac{\left\langle \mathbf{p}_{0}, \mathbf{p}_{1}, \mathbf{p}_{2}\right\rangle ^2}{\left\langle \mathbf{x}, \mathbf{p}_{0}, \mathbf{p}_{1}\right\rangle \left\langle \mathbf{x}, \mathbf{p}_{1}, \mathbf{p}_{2}\right\rangle \left\langle \mathbf{x}, \mathbf{p}_{2}, \mathbf{p}_{0}\right\rangle } \omega(\mathbf{x}).
	\end{align} 
\end{itemize}
Both of these forms are invariant under independent scaling of the homogeneous coordinates of the vertices, as well as the point of evaluation, and are thus well-defined functions over projective spaces.

Many other examples of positive geometries were identified \cite[Ch. 5]{arkani2017positive}:
\begin{itemize}
	\item Planar regions bounded by a conic section and a line.
	\item Regions in projective 3-space bounded by a quadric or cubic surface and a plane.
	\item Positive, real parts of \emph{Grassmannians} (manifolds of $k$-planes in $n$-dimensional space, denoted $G(k,n)$).
	\item Positive, real parts of \emph{toric varieties} \cite{sottile2003toric}.
\end{itemize}
These examples -- called \textbf{generalized simplices} -- all have canonical forms with constant numerators.

The canonical forms of positive geometries are additive, i.e. the canonical form of a union is the sum of the canonical forms of its parts. The poles along boundaries meeting with opposite orientation will cancel, so that only poles on the exterior boundary remain (technically, this is true for a ``signed triangulation of the empty set" -- see \cite[Ch. 3]{arkani2017positive}). This implies that the canonical forms of more complicated regions can be determined by ``triangulating" them. It follows that \emph{convex polytopes} -- which can be triangulated in the usual sense -- are also positive geometries. 

In analogy with generalized simplices, there are also various  \textbf{generalized polytopes}:
\begin{itemize}
	\item Convex regions of the projective plane bounded by straight lines and conics, which are examples of \emph{polycons}, as defined by Wachspress \cite{wachspress2016rational}.
	\item \emph{Grassmann polytopes}, in particular \emph{Amplituhedra}, which are generalizations of cyclic polytopes into Grassmannians.
\end{itemize} 
Generalized polytopes have canonical forms with a non-constant numerator. In fact, for polytopes and polycons the numerator is known as the \emph{adjoint polynomial} \cite{kohn2019projective,dasgupta2008adjoint}.

\subsection{Relation to dual polytopes}
A canonical form often has a natural geometric interpretation. In particular, the canonical rational function of a convex projective polytope $P \subset \mathbb{P}^{d}$ is the signed volume of its \emph{polar dual} $P_{\mathbf{x}}^{\ast}\subset (\mathbb{P}^{d})^{\ast}$, as shown in \cite[Ch.  7.4.1]{arkani2017positive}. The polar dual is the intersection of the dual projective cone of the polytope, with the dual hyperplane of the point $\mathbf{x} \in P$, and its volume is a rational function over $P$ computed by the integral formula
\begin{align}\label{eq:volume}
\mathrm{Vol}(P_{\mathbf{x}}^{\ast}) = \frac{1}{d!}\int_{\mathbf{y} \in P_{\mathbf{x}^{\ast}}}\frac{1}{(\mathbf{x}^{T}\mathbf{y})  ^{d+1}} \omega(\mathbf{y}).
\end{align}

\section{Barycentric Interpolation over Positive Geometries}
We claim that the canonical forms of positive geometries can be used to define barycentric interpolation in a way similar to Shepard's method. 

Consider barycentric (linear) interpolation over a \emph{segment}. Define the weight function for the endpoint $a$  as the ratio of canonical forms for positive geometries over the projective line
\begin{align}
\lambda_a(x) = \frac{\Omega(\mathbb{P}^1, \left[ a,x^{\ast}\right] )}{\Omega(\mathbb{P}^1, \left[ a,b\right])},
\end{align}
where $x^{\ast}$ is the \emph{projective dual} of the point $x$. If we choose coordinates so that $x$ is the origin and $x^{\ast}$ is the point at infinity, we get the barycentric formula (\ref{eq:segment_bary}) for linear interpolation.

\begin{figure}[h]
	\centering
	\includegraphics[width=0.3\textwidth]{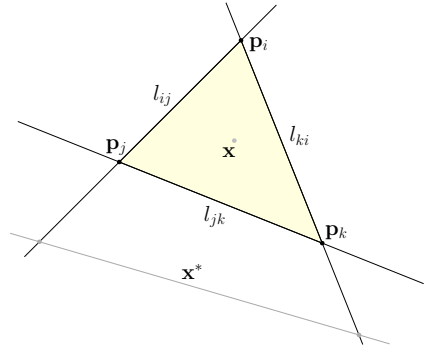}
	\caption{Triangles used for barycentric coordinates.}
	\label{fig:bary_coord}
\end{figure}

The same construction applies to a \emph{triangle} as well. For the the vertex $\mathbf{p}_{i}$, we take the ratio of two canonical forms over the projective plane -- that of the original triangle, and the triangle bounded by the two sides meeting at $\mathbf{p}_{i}$ together with the dual line of the current point $\mathbf{x}$. Using the notations of Figure \ref{fig:bary_coord}:
\begin{align}
\lambda_{i}^{\Delta}(\mathbf{x}) = \frac{\Omega(\mathbb{P}^2, \Delta(l_{ki}, l_{ij}, \mathbf{x}^{\ast}))}{\Omega(\mathbb{P}^2, \Delta(l_{ij}, l_{jk}, l_{ki}))}.
\end{align}
This is simply the usual formula (\ref{eq:triangle_bary}) for barycentric interpolation. Along each of the sides, the forms (technically, their residues) restrict to linear interpolation over a segment, as expected.

For a \emph{convex polytope} $P \subset \mathbb{P}^{2}$ we follow the same recipe -- the denominator for the vertex $\mathbf{p}_{i}$ will be the canonical form of $P$, while the numerator is the canonical form defined by the sides meeting at the vertex, with the dual line of the current point: 
\begin{align}
\lambda_{i}^{P}(\mathbf{x}) = \frac{\Omega(\mathbb{P}^2, \Delta(l_{\left( i-1\right)i}, l_{i\left( i+1\right) }, \mathbf{x}^{\ast}))}{\Omega(\mathbb{P}^2, P)}.
\end{align}
These weight functions are the Wachpress coordinates (\ref{eq:polygon_bary}) over $P$. The generalization to higher-dimensional simplices and (simplicial) polytopes is straightforward.

In each case, the numerators are defined by positive geometries that form a signed triangulation of the domain. The additivity of canonical forms under unions then implies that their sum is the canonical form of the original polytope, and thus these functions form a \emph{partition of unity}.

While these triangulations are not the most natural with respect to the original polytope, they correspond to a natural triangulation of the polar dual polytope including the origin (i.e. the current point $\mathbf{x}$). Recalling the interpretation of canonical forms as dual volumes, the numerator is seen as the volume of the dual pyramid formed by the current point and the dual face of the vertex. Thus, we connect to the earlier work of \cite{ju2005geometric}, who characterized Wachspress coordinates as ratios of polar dual volumes. Note that a triangulation of the polar dual through its vertices is analogous to Warren's triangulation-based definition of the adjoint polynomial for a polytope \cite{warren1996barycentric}.

Wachspress coordinates can be generalized also to regions bounded by subsets bounded by higher-order algebraic varieties, such as \emph{polycons} \cite{wachspress2016rational}. We omit the discussion of these cases to conform to spatial limitations. 

\section{Discussion and Future Work}
Our approach to barycentric interpolation with canonical forms of positive geometries unifies earlier approaches using adjoint polynomials, dual volumes and Shepard-like interpolation. An advantage of this framework is that it extends to positive geometries other than polytopes, such as Grassmannians and toric varieties.

We also mention that the definition of a positive geometry embeds the domain into a higher-dimensional complex manifold, thus our approach can be viewed as a multivariate generalization of complex analytical methods (contour integrals, residues) used in univariate approximation theory \cite{trefethen2013approximation}.

\subsection{Open Problem: Interpolation in Grassmannians}
As was mentioned previously, certain ``positive" subsets of Grassmannians are also examples of positive geometries. Points in a Grassmannian $G(k,n)$ can be represented by $k \times n$ matrices, and the positive geometry known as the \emph{positive Grassmannian} corresponds to \emph{totally positive} matrices, which are of great interest for approximation theory and geometric modeling \cite{gasca1996total}. 

\begin{figure}[h]
	\centering
	\includegraphics[width=0.5\textwidth]{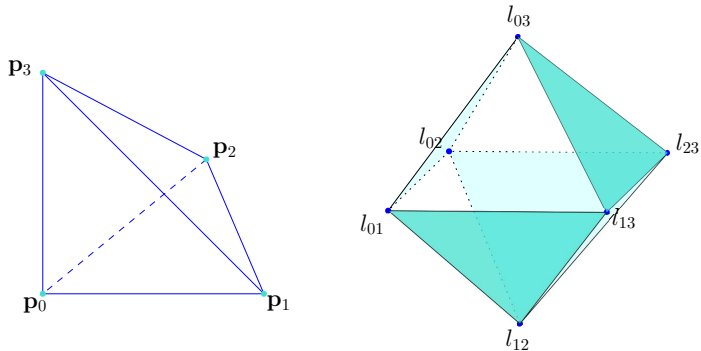}
	\caption{Tetrahedral line complex and corresponding hypersimplex. Shaded faces correspond to lines through  vertices.}
	\label{fig:grasmmannian}
\end{figure}

The Grassmannian of 2-planes in 4-space, $G(2,4)$ -- which is also the manifold of lines within 3-space  -- is particularly interesting for many applications \cite{pottmann2001computational}. $G(2,4)$ is a 4-dimensional  hypersurface in $\mathbb{P}^{5}$ cut out by a quadratic equation in Plücker coordinates, and the positive Grassmannian is a semi-algebraic subset with non-linear boundaries. The line-geometric analogue of a simplex might be related to the \emph{tetrahedral line complex}, the boundary of which is the set of lines defined by the edges of a tetrahedron. The combinatorial structure of this boundary is that of an octahedron -- an example of a \emph{hypersimplex} -- as shown in Figure \ref{fig:grasmmannian}, where each vertex represents a line along an edge of the tetrahedron, while each face represents lines through either a vertex or a face \cite{gelfand1982geometry}.

Being a positive geometry, the positive Grassmannian has a canonical form, i.e. a rational function of Plücker coordinates with poles along its boundaries \cite[Ch. 5.5]{arkani2017positive}. This suggests that generalizations of barycentric coordinates into line space might be possible using our framework.

\subsection{Open Problem: Generalized Positive Geometries for Mean-Value Coordinates}
Canonical forms are defined by rational functions, so Mean-Value Coordinates (MVCs) \cite{floater2015generalized} -- defined by transcendental functions -- are apparently incompatible with the proposed framework. We could adapt the approach of \cite{schaefer2007unified}, where MVCs are expressed as dual Shepard interpolants, after deforming the original boundary to a unit circle around the current point. A disc is not a positive geometry in the usual sense -- it lacks zero-dimensional boundary components, for example. Nevertheless, we can easily find a projectively well-defined function with singularities along a projective conic $\mathcal{C}$ given by the quadratic equation $\mathbf{x}^{T}\mathbf{Q}\mathbf{x} = 0$  (see \cite[Ch. 10]{arkani2017positive} for a lengthy discussion):
\begin{align}
\Omega(\mathbb{P}^{2},\mathcal{C}) = \frac{\pi\det(\mathbf{Q})^{\frac{3}{4}}}{(\mathbf{x}^{T}\mathbf{Q}\mathbf{x})^{\frac{3}{2}}} \omega(\mathbf{x}).
\end{align}
Observe the similarity with the transfinite form of MVCs \cite[Ch. 10]{floater2015generalized}, when $\mathcal{C}$ is a circle. This kind of canonical function is not rational and its singularities along $\mathcal{C}$ are not simple poles, but \emph{branch points} (similar to the origin of the complex plane for fractional powers and logarithms). The authors of \cite{arkani2017positive} also identified such transcendental generalizations of positive geometries as promising subjects for future research.

\subsection{Open Problem: Relation to Splines and Integral Geometry}
Formulas such as the dual volume (\ref{eq:volume}) also appear in the context of multivariate (box/simplex/cone) splines \cite{procesi2010topics}, as Laplace transforms of indicator functions. This suggests that splines and barycentric interpolants could both fit within an even more general theoretical framework related to integral geometry.

\section*{Acknowledgements}
This project has been supported by the Hungarian Scientific Research Fund (OTKA, No.124727). The author thanks Tamás Várady for support and Péter Salvi for interesting discussions. 



\bibliographystyle{plain}
\bibliography{ref}
\end{document}